\documentclass{moriond}

\usepackage{aas_macros}

\usepackage{graphicx}	
\usepackage{amsmath}	

\usepackage{amssymb}	

\usepackage{hyperref}

\usepackage{bm}
\graphicspath{{Figs/}}

\usepackage{float}
\usepackage{color}

\definecolor{orange}{rgb}{1,0.5,0}

\newcommand{\supcite}[1]{}

\newcommand{\HI}{\mbox{H{\scriptsize I}}} 
\newcommand{\mMsol}{\mathrm{M_\odot} } 

\newcommand{\Photo}{}

\begin{document}
\vspace*{4cm}
\title{Current status and future of cosmology with 21cm Intensity Mapping}

\author{ R\'eza Ansari }

\address{Universit\'e Paris-Saclay, CNRS/IN2P3, IJCLab, 91405 Orsay, France}

\maketitle
\abstracts{ 
21cm Intensity Mapping (IM) has been proposed about 15 years ago as a
cost effective method to carry out cosmological surveys 
and to map the 3D distribution of matter in the universe, over a large
range of post EoR redshifts, from z=0 to z=6. 
Since then a number of pathfinder instruments have been built, such as CHIME or Tianlai.
Several other ones will be commissioned in the next few years (HIRAX, CHORD, BINGO), while 
even larger arrays, with several thousand antennae are being considered for the next generation experiments.
We will briefly review the 21cm cosmology of the Epoch of Reionisation (EoR), and we will 
then focus on IM for late time cosmology. After presenting some of the promises of this 
technique to constrain the cosmological model, dark energy and inflation, we will 
review  some of the instrumental and scientific challenges of IM surveys. 
The second part of the paper presents an overview of the ongoing and future experiments, as 
well as recent results by GBT, CHIME and Tianlai. 
}



\section{Introduction} 
\label{sec:introduction}

Although the 21cm cosmology is mostly concerned with probing the dark ages and EoR, 
this paper is rather focused on late time cosmology, and mapping the cosmic matter distribution using
the 21cm radio emission or absorption of neutral hydrogen gas (\HI) through Intensity Mapping (IM). 
This technic refers to the detection of underlying structures formed by 21cm sources such as galaxies, 
without requiring  detection of individual sources \cite{2001JApA...22...21B,2004MNRAS.355.1339B}.
A brief overview of EoR science with 21cm is presented in this section, while  the idea and instrumental 
concepts behind Intensity Mapping would be developed in the following  sections.  

Observations revealing the evolution of the universe during the dark ages and the reionisation era are crucial for understanding 
the formation of structures in the universe \cite{2006PhR...433..181F,2008PhRvD..78j3511P,Morales&Wyithe2010,Furlanetto:2019jso}. 
Dark ages refers to the period extending from Cosmic Microwave Background (CMB) last scattering era, at a redshift $z \sim 1100$ 
to the cosmic dawn, which corresponds to the birth of first stars and galaxies. 
The intense and energetic (UV, X) radiation from theses first sources initiated the process of ionising the neutral 
gas, marking the start of the Epoch of Reionisation (EoR).
The 21cm hyperfine transition of atomic hydrogen (\HI) is one the only spectral features that can be used to probe 
this era, as the universe contains no sources, but only gas, mostly hydrogen at high redshifts $(z \gtrsim 30)$. 
The \HI \, gas leaves an imprint on the cosmic background radiation (CMB) if the spin temperature $T_s$ differs
from the CMB temperature $T_{CMB}(z)$ at the corresponding redshift. The redshifted \HI hyperfine transition, 
at a frequency of $\nu_{21}/(1+z)$ with $\nu_{21} \simeq 1420.4 \mathrm{MHz}$  will appear as an emission, 
respectively an absorption feature,
if the spin temperature $T_x(z)$ is higher, respectively lower, than $T_{CMB}(z)$.   
The history of the evolution of the spin temperature during the dark ages and EoR is rather complex \cite{2012RPPh...75h6901P}.
$T_s$ tracks at first $T_{CMB}$, then decouples around $z \gtrsim 150$ and decreases
due to collision coupling with the gas kinetic temperature $T_K$. 
The $T_s \leftrightarrow T_{CMB}$ equilibrium is restored around $z \sim 50$, when the atomic collision rates 
becomes ineffective to maintain $T_s \leftrightarrow T_K$ coupling, due to gas dilution by the expansion. 
After the formation of the first stars and galaxies, the spin temperatures 
gets coupled to the gas temperature again in the UV photon bath at $z \sim 30$, through the Wouthuysen-Field effect. 

The measurement of the 21 cm emission temperature and its anisotropies as a function of redshift would therefore allow 
to precisely identify  the different stages of  the universe's evolution during the dark ages and EoR ($10 \lesssim z \lesssim 100$). 
In addition, the temperature anisotropies trace the distribution of matter in the linear regime 
over a broad range of wave modes (k-scales) at these redshifts, whereas non linearities affect a significant fraction 
of k-scales at later cosmological times $(z \sim 1)$.

Although a number of dedicated instruments have been built over the last twenty years to observe the EoR 21cm signal, 
no undisputed observation has yet been reported. This is explained by the many challenges that needs to be overcome:
the redshifted 21cm feature is located in the frequency range $10-50 \mathrm{MHz}$ for redshifts $30\lesssim z \lesssim 150$, 
which suffers from significant ionospheric absorption and diffraction as well as from major terrestrial, man-made  disturbances (RFI). 
Moreover, the 21cm cosmological signal is very faint, in absolute terms, and completely buried in the foregrounds, 
specially the Galactic synchrotron and radio source emissions, which dominate this signal by  4 to 5 orders of magnitude.
In addition, the reionisation history is very poorly constrained, making the experimental adventure quite risky, given the 
limited bandwidth that can be covered by any given instrument. 

Some experiments, such as PAPER \cite{2010AJ....139.1468P}, SCI-HI \cite{2014ApJ...782L...9V} or EDGES \cite{2017ApJ...835...49M} 
have targeted the global spectrum measurement and the detection of the distortions 
in the spectral shape of radio emission, due to EoR 21cm signal \cite{2013PhRvD..87d3002L}. 
Although some authors have claimed a possible signal \cite{2018Natur.555...67B}, the detection has not been confirmed. 
Other groups, have developed complex wide band interferometric instruments to detect the inhomogeneties of the cosmological 21cm signal 
and measure the associated power spectrum. 
LOFAR \cite{2013A&A...556A...2V,2019A&A...622A...1S}  and NenuFAR \cite{2021Galax...9..105B} \supcite{2021sf2a.conf..211M} 
have deployed antennae in Europe, LWA \cite{2018AJ....156...32E} is a large dipole array in New Mexico (USA), 
while MWA \cite{2013PASA...30....7T} or HERA \cite{2014ApJ...782...66P,HERA2017} observe from Australia. 
Recent upper limits on the 21cm power spectrum from these experiments  can be found in \cite{2020MNRAS.493.1662M} for LOFAR, 
in \cite{2016ApJ...833..102B} for MWA and in \cite{2022ApJ...925..221A} for HERA. Detection of the 21cm EoR signal is also the main goal of 
the SKA-low instrument of the future SKA (Square Km Array) observatory \cite{2015aska.confE..10M}. \supcite{2020MNRAS.494.4043M} 
  
An brief overview of post EoR $(z<6)$ 21cm Intensity mapping and its scientific promises  and challenges 
is presented in section \ref{sec:21cmIM}.
The instrumental concepts suited for such surveys, as well as key technical issues are discussed in section \ref{sec:instrumentconcept}.
Ongoing and planned intensity mapping instruments and surveys and their latest results are presented in section \ref{sec:experiments}. 
The last section, \ref{sec:conclusion} gives an outlook and expectations for the near future. 
 
\section{21cm Intensity Mapping} 
\label{sec:21cmIM}

The large scale distribution of matter in the universe is a powerful cosmological probe, used 
to reconstruct the universe expansion history, and to determine the statistical properties of the initial density fluctuations. 
Indeed, the quantum fluctuations, relics of the early universe inflationary phase are considered to be 
the seeds that generated the large structures, visible in the late universe, through the gravitational instability \cite{2000cils.book.....L,2012arXiv1208.5931K}.
The LSS statistical properties, mostly encoded in the shape of the spatial correlation function $\xi(r)$ 
or the power spectrum $P(k)$, depends on the cosmological model and its parameters.  
The evolution of large structures with redshift, the growth rate of the structures in particular, 
is sensitive to changes in gravity \cite{2013PhRvD..87b3501A}, as well as to the neutrino masses, 
which, depending on their masses, partially erase small-scale structures \cite{2015APh....63...66A}. 
The Baryon Acoustic Oscillations (BAO's) correspond to a preferred structure scale in the distribution of galaxies, 
originating from the baryon-photon plasma oscillations, prior to the decoupling. The BAO peaks, when observed at different redshifts, 
can be used as a standard ruler to reconstruct the cosmic expansion history, through the measurement of the Hubble 
parameter $H(z)$ and angular diameter distance  $d_A(z)$ \cite{2003ApJ...594..665B}. \supcite{2003PhRvD..68f3004H}

Historically, most cosmological surveys have been carried out using optical instruments 
through spectroscopic or photometric observations. 
Recent constraints on cosmological parameters derived from eBOSS  and DES optical surveys can be found  in 
\cite{2021PhRvD.103h3533A}  and  \cite{2022PhRvD.105b3520A}. 
However, mapping matter distribution in the universe in the radio wavelengths, 
through the observation of the redshifted 21 cm line of the atomic hydrogen (\HI), is a complementary approach to optical
surveys to constrain cosmology and dark energy. The 21 cm line is the only astrophysical spectral feature in the L band 
($\sim \mathrm{GHz}$). 
It can therefore be used to determine  unambiguously  the redshift of an astrophysical object.   
Nevertheless, the detection of galaxies at 21cm needs a very large collecting area. 
The ALFALFA survey \cite{ALFALFAFullSample2018} carried out with the Arecibo antenna,
one of the largest radio-telescopes in the 
world \footnote{The Arecibo 305 m telescope is being decommissioned, 
following damages to its structure \url{https://www.nsf.gov/news/news_summ.jsp?cntn_id=301674} }, 
with a primary reflector 300 m in diameter, detected objects up to a redshift of 0.2. 
SKA will be able to extend this limit and allow the observation of gas-rich galaxies up to $z \sim 0.5$. 

Most of the cosmological information of large structures (LSS) is found at scales larger than a few Megaparsecs (Mpc).  
Therefore, the detection of individual galaxies is not mandatory for LSS studies.  
This is the essential idea behind the Intensity Mapping technique, which seeks 
to measure the aggregate 21 cm radiation of all the galaxies (a few hundred) contained in universe cells with a volume of a few hundred $\mathrm{Mpc^3}$.  
A cosmological survey becomes then possible with more modest instruments \cite{2008PhRvL.100i1303C,2008arXiv0807.3614A,2010ApJ...721..164S,2012A&A...540A.129A}, 
with a collection area of about few times $10^4 \mathrm{m^2}$. 
However, the LSS cosmological signal has an average surface brightness of less than 1 mK, and is 
therefore totally overwhelmed by foreground emissions, mainly from the Milky Way synchroton 
emission and radio sources. These foreground emissions, with a temperature of $T_\mathrm{fgnd} \sim 2-5 \mathrm{K}$ 
in the coldest parts of sky around 1 GHz, are in general about 
few thousand times brighter than the cosmological \HI \, emission, 
while the instantaneous receiver noise is still about ten times larger with $T_\mathrm{sys} \sim 50 \mathrm{K}$. 
The reduction of fluctuations from instrumental noise $(T_\mathrm{sys})$ is achieved through long integration time, 
a few hours for each direction of the sky.

Extraction of the cosmological 21cm signal in the presence of these foregrounds  
is among the main IM scientific challenges. 
Several methods of separating the signal from the foreground emissions have been proposed 
which are all based on the smooth variation of the foreground brightness with frequency
\cite{2012A&A...540A.129A,2014MNRAS.441.3271W}. 

Since the early works on 21cm Intensity Mapping as a tool for cosmology \supcite{e.g.\cite{2009astro2010S.234P}}, many authors 
have studied the science reach of such surveys, either generically \cite{2015ApJ...803...21B,2018ApJ...866..135V}, 
or targeting specific existing instruments such as FAST \cite{2017A&A...597A.136S}, 
or the SKA \supcite{2021MNRAS.505.3698W} \cite{2022MNRAS.509.2048S}. 

\begin{figure}[thbp]
\includegraphics[width=0.46\textwidth]{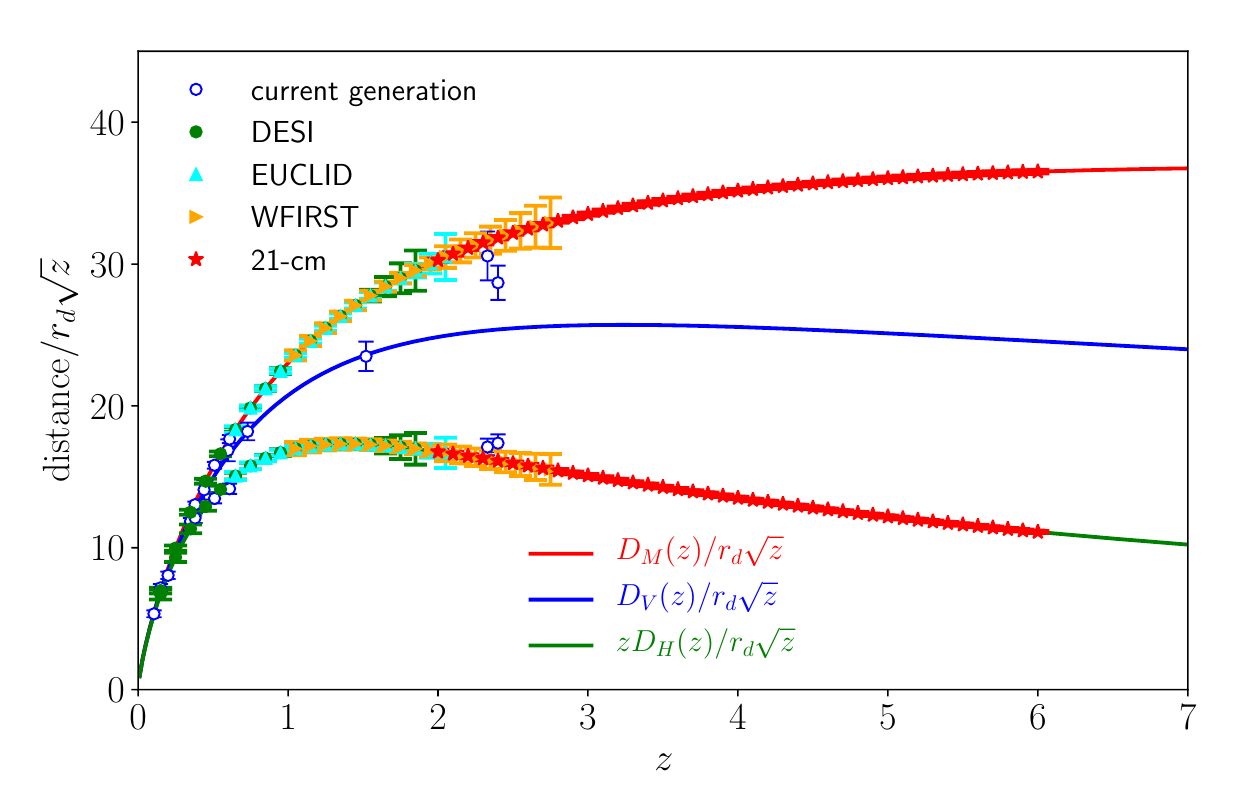} \hspace{2mm} 
\includegraphics[width=0.53\textwidth]{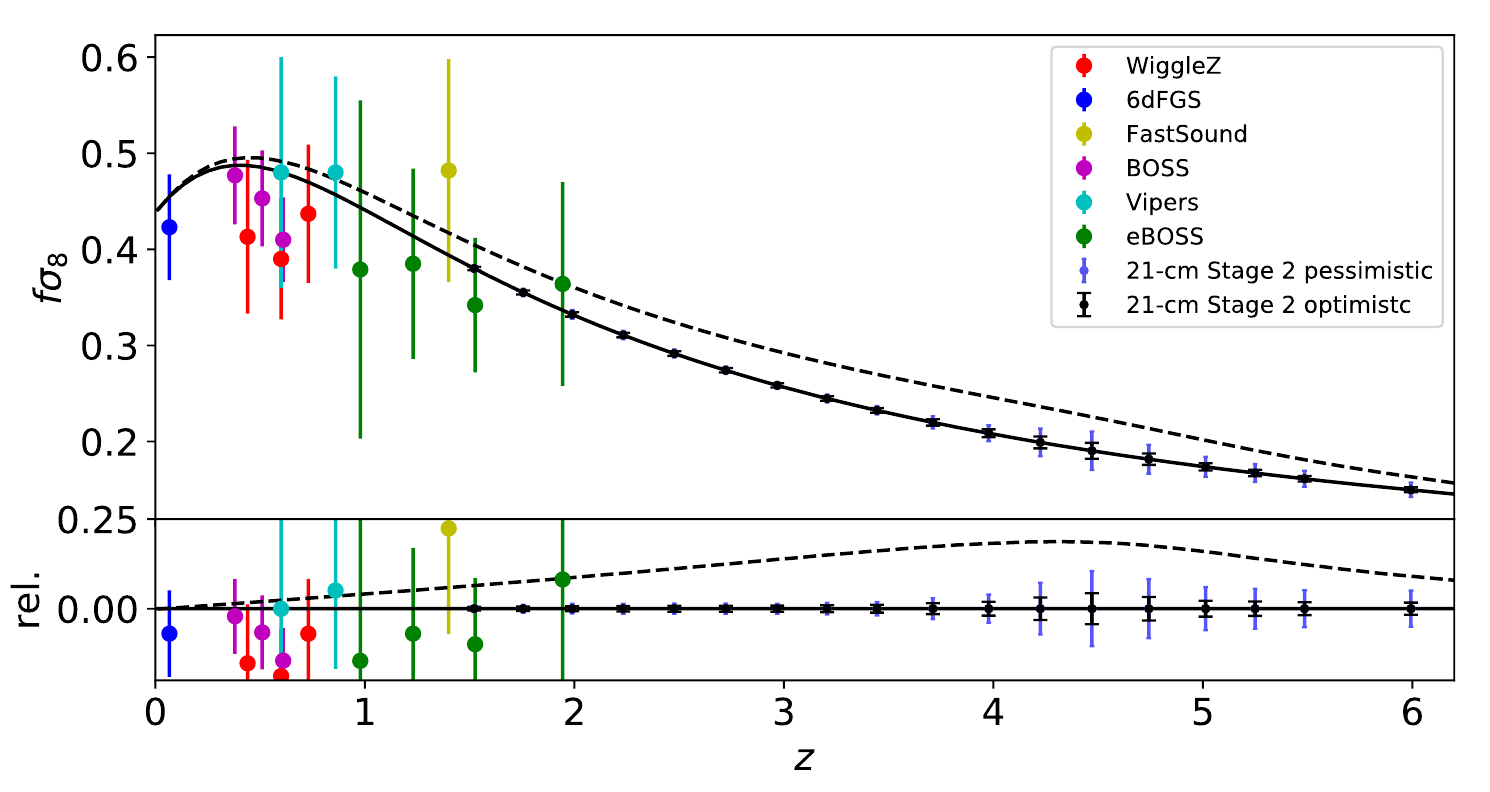}
\caption[]{Performance of a 21cm Intensity mapping survey with a Stage-II instrument, compared to the
  next optical surveys. Left: Expansion history reconstructed through different BAO distance
  scales measured as a function of redshift.
  Right: Statistical precision on the determination of the  
  growth rate of structures. Figures adapted from reference \cite{2018arXiv181009572C}.  }
\label{fig:cosmoIM}
\end{figure}

The white paper \cite{2018arXiv181009572C} present the science goals of an 
ambitious 21cm Intensity Mapping survey covering a broad redshift range, 
up to $z \sim 6$ with a very large dish array radio interferometer like PUMA \cite{2019arXiv190712559B}. 
Figure \ref{fig:cosmoIM}, adapted from \cite{2018arXiv181009572C}, shows the precision that could be reached on the 
determination of the transverse and longitudinal BAO scales as a function of redshift,  as well as for the structure growth rate $f \sigma_8$, 
assuming  that the instrument can be built and operated, with map making and foreground subtractions reaching the  projected performances. 
Thanks to the very large surveyed volume, such an IM experiment would significantly outperform surveys 
by the latest optical instruments (Rubin/LSST, DESI, WFIRST, Euclid). Radio observations can also be 
used to search for  non-gaussianities and inflationary features in the reconstructed 3D LSS maps.

\section{Instrumental concepts and challenges } 
\label{sec:instrumentconcept}
\begin{figure}
  \parbox{0.50\textwidth}{
  \includegraphics[width=0.48\textwidth]{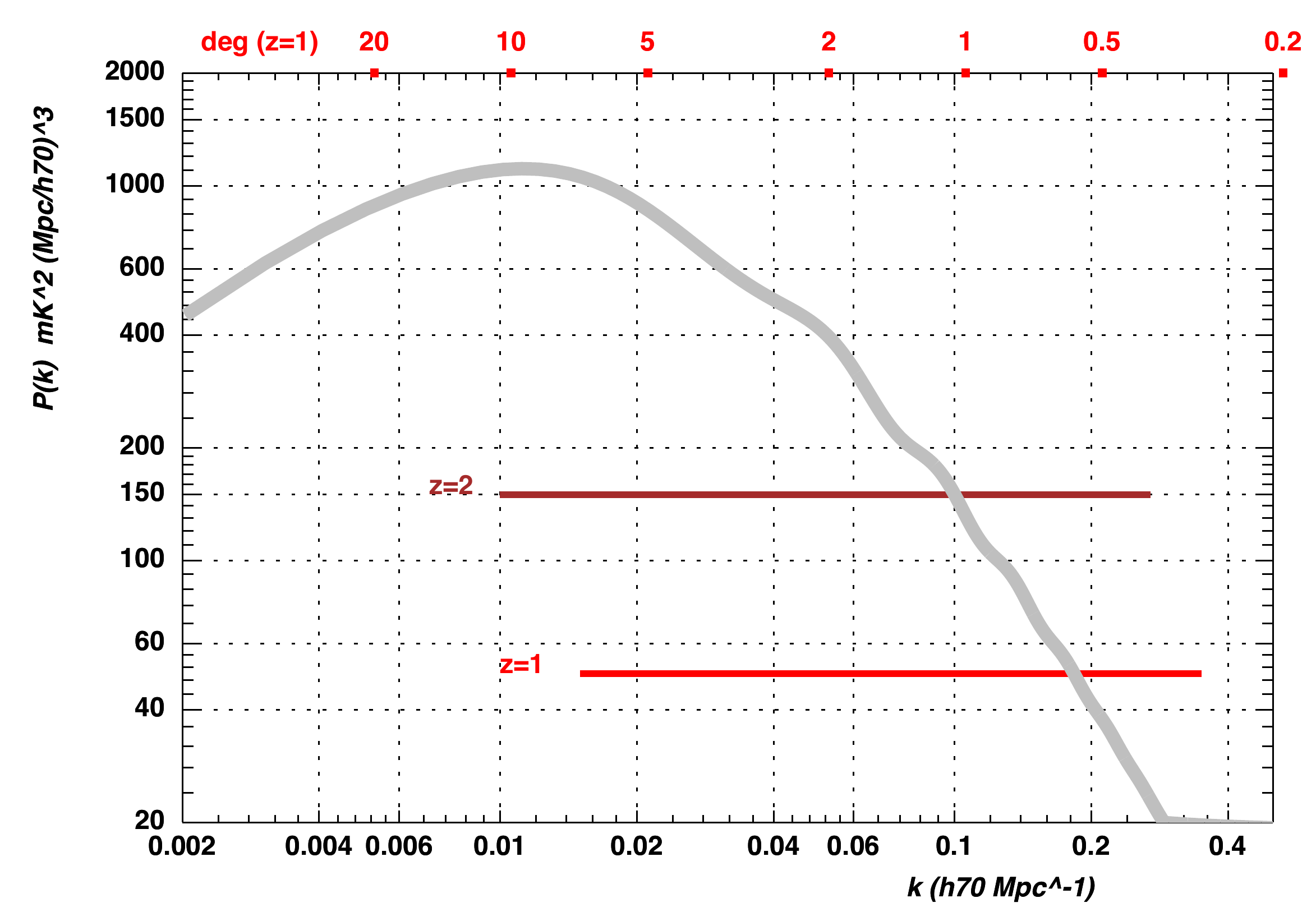} 
  }
  \parbox{0.5\textwidth}{\small The power spectrum $P_{21}(k)$ in $\mathrm{mK/(Mpc/h_{70})^3}$ is plotted 
as a function of the comoving wavenumber $k$ in $h_{70} \mathrm{Mpc^{-1}}$.
The top scale shows the angular scales at $z \sim 1$ for the corresponding transverse $k_\perp$ wavemodes. 
The expected projected noise level, due to the instrument noise for $T_\mathrm{sys} \sim 50 \mathrm{K}$, in the white noise approximation, 
is also shown, at $z=1$ and $z=2$ for a survey of $\sim 5000 \mathrm{deg^2}$, with  an  array of $\sim$ 1600 dishes, 
covering $\sim 200 \times 200 \mathrm{m^2}$. The noise level line extent shows the approximate accessible $k_\perp$ range. }
\caption[]{ 21cm brightness temperature fluctuations power spectrum $P_{21}(k)$ }
\label{fig:21cmPkPnoise}
\end{figure}

As already mentioned in section \ref{sec:21cmIM}, the relatively low
radio brightness of \HI \, clumps and galaxies limits the possibility of their detection to the vicinity, in cosmological
sense, of our galaxy, with the available radio instruments. A galaxy with an \HI \, mass of $10^{10} \mMsol$, \
which is already a quite massive hydrogen cloud, would have a 21cm brightness $S_{21} \simeq 10 \, \mathrm{\mu Jy}$ if located at a redshift $z =1$. 
Its detection, with a reasonable integration time of a few hours will require more than $10^5 \mathrm{m^2}$ of collecting area. 
Most of the cosmological information in the LSS is located at scales of a few arc-minutes or larger, as can  
be seen on  figure \ref{fig:21cmPkPnoise}, which shows the approximate level of the 21cm brightness temperature 
fluctuations power spectrum $P_{21}(k)$ for $z \sim 1$. The linear matter power spectrum has been converted into temperature,
assuming an \HI to baryon fraction $f_{H_I} \sim 1 \%$, and using the following formulae, adapted from \cite{2008PhRvD..78j3511P,2012A&A...540A.129A}:
\begin{eqnarray}
P_{21}(k) & \sim & \left( \langle T_{21} \rangle \right)^2 \times P_{LSS} (k) \\
\langle T_{21} \rangle & \simeq & 0.042 \mathrm{mK} \, \frac{\Omega_{H_I} }{10^{-3}} \frac{H_0}{H(z)} (1+z)^2 
\end{eqnarray}

The cosmological information encoded in the LSS is statistical in nature, so a reasonably large volume of the universe 
needs to be surveyed to extract the information with low enough statistical errors. Instruments suitable for 21cm Intensity Mapping 
thus needs to have a large instantaneous field of view (FOV $\gtrsim 10 - 100 \mathrm{deg^2}$) and a large bandwith 
($\Delta \nu \gtrsim 100 \mathrm{MHz}$), to be able to survey a large fraction of the sky with large integration time, 
over a significant redshift range. 

Progress in the L-band analog electronics have made room temperature RF amplifiers quite competitive with low noise cryogenic 
electronic used in the large radio telescopes. Noise temperatures below $T_{noise} \lesssim 30 \mathrm{K}$ can indeed now be reached 
by room temperature receivers. Large bandwidth interferometers with large number of feeds have become viable 
and cost effective, thanks to advances in digital electronic and computing (multi core CPU and GPU's), combined 
with progress in room temperature RF analog electronic. 

Interferometers are most often used to reach high angular resolution, thanks to long baselines, but at the expense of a sparse sampling of the 
angular wave-mode or $(u,v)$ plane. High resolution is not needed for LSS mapping for cosmological purposes, while high sensitivity 
is crucial, given the low signal strength. Although large dishes equipped with a multi-feed or phased array in their focal plane 
are a possible option, densely packed interferometric arrays, using rather small reflectors ($D \sim 5-10 \mathrm{m}$),
operating in transit mode, are the type of instruments most widely considered for intensity mapping  surveys. 
The small size of reflectors insures a large FOV $(\sim \left( \frac{\lambda}{D} \right)^2 \mathrm{srad})$, 
and the dense packing concentrates the sensitivity in the $k_\perp$ range useful for LSS. 
Transit mode operation is well adapted for surveying large area of the sky, while reducing instrument 
complexity and cost. Initially, cylindrical reflectors, with their axis or the focal line oriented along the north-south 
direction were proposed \cite{2006astro.ph..6104P}, and implemented in the Pittsburgh CRT (Cylindrical Radio Telescope) 
prototype and then in CHIME, as well as in Tianlai. It was then realised that packed dish arrays might have 
some advantages, despite a smaller FOV and the need to change dish pointing in declination  to cover a large enough sky area.
PAON4 and Tianlai  are early examples of dish based pathfinder instruments (see section  \ref{sec:experiments}). 

Radio instruments are inherently diffraction limited with their angular resolution degrading at longer wavelengths, hence with redshift. 
The projected spatial resolution will in addition vary with redshift, depending on the cosmological distance scales, namely 
the line of sight distance $d_{LOS}(z)$ and Hubble parameter $H(z)$. Values of transverse and longitudinal spatial resolution 
of maps obtained with a radio array covering a $\sim 100 \times 100 \mathrm{m^2}$ area and with a $250 \mathrm{kHz}$ resolution 
are gathered in table \ref{tab:angresol-fz}. The approximate level of instrument noise projected on sky for a $\sim 1600$ 
element arrays, covering a $200 \times 200 \mathrm{m^2}$, surveying $\sim 5000 \mathrm{deg}^2$ over a year is shown in 
the figure \ref{fig:21cmPkPnoise}, computed according to the scaling formula in \cite{2012A&A...540A.129A}. 
One can see that the resolution, hence the projected noise level degrades quickly with increasing redshift.
A few hundred element interferometer with a few thousands $\mathrm{m^2}$ 
collecting area might be sufficient to detect the signal up to $z \lesssim 1$, 
while larger instruments, with $10^4$ feeds and $\sim 10^5  \mathrm{m^2}$ are required 
to reach higher redshifts $z \sim 2-3$. 

\begin{table}

\begin{center}
\begin{tabular}{|c|ccccc|} 
\hline 
$z$ & $\delta \theta$ (arcmin) & $d_{LOS}$ (Mpc) & H(z) (km/s/Mpc) & $a_\perp$ (Mpc) & $a_\parallel$ (Mpc) \\
\hline 
0.5 & 13' & 1950 & 90 & 7.2 & 1.3 \\
1.0 & 17' & 3530 & 120 &  17.3 & 1.8 \\ 
1.5 & 21' & 4660 & 160 &  28.5 & 2.2 \\
2.0 & 26' & 5310 & 200 &  39. & 2.3 \\
2.5 & 30' & 5970 & 255 &  50. & 2.5 \\
3.0 & 34' & 6500 & 308 &  63. & 2.7 \\
\hline 
\end{tabular}
\caption{Evolution of the transverse $a_\perp$ and the radial $a_\parallel$ resolutions of a radio array with 
$L=100 \mathrm{m}$ size, and  $250 \mathrm{kHz}$ frequency resolution as a function of redshift, 
in the standard  $\Lambda$CDM cosmology. \label{tab:angresol-fz}}
\end{center} 
\end{table} 

Reconstructing sky maps, from visibilities which are cross correlation signals measured by interferometers has been 
and is still a technical challenge and major efforts have been devoted to develop accurate and efficient map making methods 
and tools. The transit operation mode, with observations covering the full 24 hours of right ascensions, combined 
with the large covered sky area, has led to the development of mathematically rigorous and efficient methods to 
reconstruct sky maps from transit visibilities. These methods operate in the spherical harmonic $(\ell,m)$ domain, and 
take advantage of the full 24 hours coverage to decompose visibilities into m-modes. The huge matrix
representing the linear relation between time dependent visibilities and the sky can be written as a block diagonal 
matrix in this representation, making the numerical problem tractable \cite{2014ApJ...781...57S,2016MNRAS.461.1950Z}. 

As stated in section \ref{sec:21cmIM} the 21cm signal is several orders of magnitude fainter than the broad band emissions 
from the Milky Way the te radio-sources. The foreground subtraction or component separation is a major difficulty in Intensity Mapping surveys, 
probably more challenging than in CMB experiments. The different approaches relie all on the 
foreground brightness varying smoothly with frequency, behaving as power laws, while the signal that follows 
the matter density fluctuations is structured along the frequency, corresponding to the radial direction. 
The inherent frequency dependence of the interferometer beam makes the foreground subtraction even more difficult. 
This is usually referred to as {\em mode mixing} as the angular modes probed by a given baseline changes with frequency. 

21cm Intensity Mapping shares many technical aspects, map making and foreground subtraction in particular 
with the search for the 21cm signal from the EoR \cite{2015PhRvD..91b3002D}. 
Although signal-foreground separation might be effective on a per visibility  basis with filtering along the frequency axis \cite{2012ApJ...756..165P,2012ApJ...752..137M}  many authors have explored the methods where signal and foreground components
are projected into different sub spaces or modes \cite{2014ApJ...781...57S,2015PhRvD..91h3514S,2016MNRAS.461.1950Z,2019AJ....157....4Z}.

\section{Past, ongoing and future experiments} 
\label{sec:experiments}
We review here some of the experimental and observational efforts initiated in the last decade to explore and possibly establish 
the faisability of observing large scale structures through 21cm intensity mapping. 
We start by mentionning some of the pioneering work performed using the GBT and Parkes radiotelescopes. 
We will then describe the dedicated pathfinder instruments,  CHIME and Tianlai which have been specifically built to carry IM surveys, 
before presenting briefly few other projects, which are in the construction phase and will become operational in the coming years.  
Obviously, the list of projects mentioned here is incomplete. 

The HI Parkes All Sky Survey (HIPASS) \cite{2004MNRAS.350.1195M} was carried out from 1997 to 2001 with the 
64 meter Parkes Australian telescope, equipped with the 21cm multi beam receiver \cite{1996PASA...13..243S}. 
It covers a large fraction of the southern, but also the northern sky, in the velocity range $-1280 < cz < 12700 \mathrm{km/s}$,
or $z \lesssim 0.042$. A positive cross-correlation of 21cm signal from HIPASS data cubes with galaxies from 
the 6dF \cite{2009MNRAS.399..683J} survey was reported in 2009 \cite{2009MNRAS.394L...6P}. 

The fully steerable 100m diameter Green Bank Telescope 
(GBT) \footnote{GBT: \href{https://greenbankobservatory.org/science/telescopes/gbt/}{https://greenbankobservatory.org/}} 
has also been used to statistically detect the redshifted 21cm signal, in cross-correlation with optical surveys. 
Observations with the GBT have been carried out from 2010 to 2015, using the 680-920 MHz prime focus receiver covering 
the $0.6 \lesssim z \lesssim 1.0$ redshift range. 
A first cross-correlation detection from observation in DEEP2 fields \cite{2013ApJS..208....5N} was reported
in 2010 \cite{2010Natur.466..463C}, confirmed with more data, using the WiggleZ \cite{2010MNRAS.401.1429D} 
survey in 2013 \cite{2013ApJ...763L..20M}. The detection of the cosmological 21cm autocorrelation signal has
also been claimed using the same data set \cite{2013MNRAS.434L..46S}. 
More recently, using a subset of the GBT data, corresponding to about 100 hours of observations covering $\sim 100 \mathrm{deg^2}$ and 
the SDSS-IV eBOSS and WiggleZ spectroscopic redshift catalogs, the \HI \, gas fraction at redshift $z \sim 0.8$ has been measured \cite{2022MNRAS.510.3495W},
more precisely the product of hydrogren mass fraction, its bias and galaxy- hydrogen correlation 
coefficient ($\Omega_{H_I} b_{H_I} r_{H_I} \sim 0.35-0.6 \times 10^{-3}$ at $ z \sim 0.8$).   \\[1mm] 

Tianlai (Cosmic Sound in Chinese) \supcite{Chen2012} is an international project led by NAOC, with US, French 
and Canadian contributions and is exploring 21cm Intensity Mapping. 
The collaboration operates two pathfinder instruments which have been built and installed in a radio quiet area in north-western China, 
a cylindrical reflectors (TCI) and a dish array interferometer (TDA).
The observatory is located at ($91^\circ 48' \mathrm{E}$, $44^\circ 09' \mathrm{N}$) and an altitude of $\sim 1500 \mathrm{m}$, 
near Hongliuxia, about 500 km east of Urumqi, in the Xinjiang province. 
The feeds have been designed to cover a broad frequency range (400-1430 MHz), although the current digitisation and correlator electronic 
can only handle a narrower 100 MHz frequency interval. The frequency band is tunable by adjusting the local oscillator and
changing the analog filters. All observations have been performed in the frequency band 700-800 MHz up to now. 

The TCI  is composed of three cylinders, each 40 m long and 15 m wide. Only the central part of the focal line is currently instrumented 
with receivers. The three cylinders are equipped with a total of 96 dual polarisation feeds, representing 192 RF signals. 
The digital correlator computes 18528 visibilities from sampled signals, and channelised into $1024$ frequency channels 
with $122 \mathrm{kHz}$ frequency resolution.
Reference \cite{2020SCPMA..6329862L} present the basic system performance for the cylinder array. 

The TDA is composed of 16 on-axis dishes, each 6 meter in diameter, arranged in a circular layout, with a central dish, 
surrounded by six dishes arranged as an hexagon, and then an outer ring with nine dishes. 
Although dishes are fully steerable, the array operates in transit mode, with all dishes fixed and pointed towards a given 
direction in the meridian plane. 
Each dish is equipped with a dual polarisation receiver, representing a total of 32 RF signals, sampled at 250 MSPS and 14 bits.
A total of 528 visibilities (496 cross + 32 auto) are computed by an FPGA based correlator for 512 frequency channels with 
$244 \mathrm{kHz}$ resolution. A few thousand hours of observations have already been performed, with a significant fraction 
spent toward the NCP (North Celestial Pole). Indeed, one of the advantages of the dish arrays is their ability to be pointed toward 
a given declination band. By observing toward the NCP, it is possible to reach a high sensitivity over a small sky area. 
Detailed description of the Tianlai dish array and its performance, as well as preliminary results from deep 
observations toward the NCP can be found in reference \cite{2021MNRAS.506.3455W}.    

Smaller instruments such as PAON4 \cite{2020MNRAS.493.2965A} and BMX \cite{2020SPIE11445E..7CO}
have also been built to explore specific technical aspects of dish arrays operating in transit mode. 
PAON4 is a small, 4-dish test interferometer located at the Nan\c{c}ay radio observatory in France, operational since end of 2015. 
It consists in four D=5 m diameter dishes, equipped with dual linear polarisation feeds, 
arranged in a triangular layout with a fourth dish in the center. PAON4 is being upgraded and will serve as the qualification 
instrument for the new IDROGEN digitisation and processing electronic system. 
These FPGA (Field Programmable Gate Array) based boards exploits the 
White Rabbit technology \footnote{ White Rabbit clock synchronisation over ethernet : \href{https://white- rabbit.web.cern.ch}{https://white- rabbit.web.cern.ch} } 
for precise clock synchronisation and can digitise L-band signals at the receivers, without frequency shifting.
They can transfer waveform signals, or  frequency components as digital streams over high throughput ($>20 Gb/s$) 
ethernet optical links.  

\begin{figure}[htbp] 
\centerline{\includegraphics[width=0.9\textwidth]{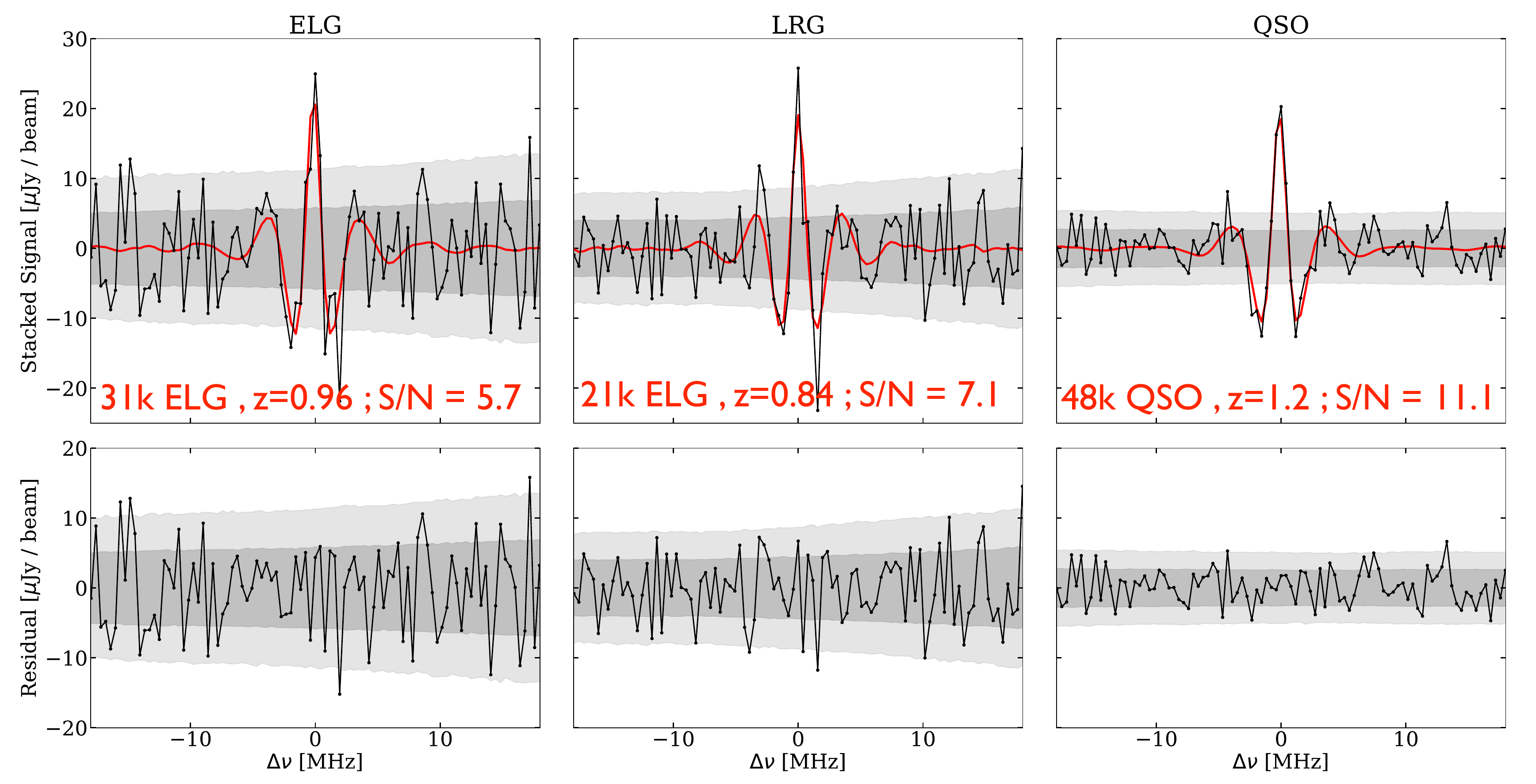}  }
\label{fig:chimeeboss}
\caption[]{  Cross correlation of CHIME intensity maps with  eBOSS galaxies and quasars. Top figures show the actual stacked signal in black, 
  stacked on the positions of ELG's with mean redshift $\langle z \rangle = 0.96$ (left), LRG's with $\langle z \rangle = 0.84$ (center) and QSO with $\langle z \rangle = 1.2$ (right).
  The red curve corresponds to the best fit model and the bottom figures show the residuals, after 
  best fit model subtraction.  Adapted from \cite{2022arXiv220201242C} }
\end{figure}

The Canadian Hydrogen Intensity Mapping Experiment (CHIME) \cite{2014SPIE.9145E..22B} is a cylinder based interferometer
designed and built for 21cm Intensity Mapping. It consists of four north-south oriented cylinderical reflectors, 
each $100 \mathrm{m}$ long and $20 \mathrm{m}$ wide, located at the DRAO observatory in Penticton, British Columbia (Canada). 
The instrument has a very broad bandwidth and covers the 400-800 MHz frequency range, corresponding to redshifts between 0.8 and 2.5.  
Each cylinder is fully instrumented and equipped with 256 dual polarisation feeds.  The 2048 $(4 \times 2 \times 256)$ analog signals
are processed by  an FX correlator. The F-engine uses FPGA custom designed electronics to perform digitisation and frequency decomposition 
into 1024 channels, each $390 \mathrm{kHz}$ wide. Correlations are computed for all $N^2$ feed pairs by the X-engine, which uses 
custom electronic for data exchange and 256 GPU nodes which perform the actual correlation computations. An overview of 
CHIME performance can be found in \cite{2022arXiv220107869T}. 

Thanks to their large FOV and bandwidth, Intensity Mapping instruments are also very efficient in detecting transient radio sources, 
such as FRB's (Fast Radio Bursts) \cite{2019A&ARv..27....4P} and pulsars. However, a specific backend is required to process the radio signals 
to search for fast transients. CHIME has proved to be a powerful radio burst and pulsar observation machine 
and has detected several hundred FRB's over a one year observation period \cite{2021arXiv210604352T}. 

Despite CHIME successful operation, efficient data processing and impressive performance, the detection of the 21cm cosmological 
signal appears more challenging than anticipated. However, CHIME has published very recently, in February 2022, \cite{2022arXiv220201242C} 
a sophisticated analysis showing convincing evidence for a correlation signal between CHIME intensity maps, and eBOSS \cite{2016AJ....151...44D}
galaxies and quasars at redshifts $0.78 < z < 1.43$. Maps were reconstructed from over 100 nights of CHIME observations, then stacked on the angular and frequency 
positions of eBOSS galaxies and quasars, after filtering and foreground subtraction.  Figure \ref{fig:chimeeboss}, adapted from \cite{2022arXiv220201242C},
shows the stacked signals as a function of the frequency shift, for ELG's (Emission Line Galaxies), LRG's (Luminous Red Galaxies) and quasars (QSO),
together with the best fit model. 

BINGO (Baryon acoustic oscillations from Integrated Neutral Gas Observations) is a single dish instrument being built in Brazil \cite{2021arXiv210701634W}.
It uses a off-axis $1600 \mathrm{m^2} , D \sim 50 \mathrm{m}$ primary reflector which illuminates a focal plane 
equipped with 28 horns through a secondary mirror. The whole optical setup is fixed and will observe a $15^\circ$ wide 
declination band centered on $\delta = 15^\circ$. The instrument covers the frequency range 980-1260 MHz corresponding 
to redshifts $0.13 < z < 0.45$.

HIRAX (Hydrogen Intensity and Real time Analysis eXperiment) \cite{2021arXiv210913755C} is a packed array interferometer using $D=6 \mathrm{m}$ 
dishes and shares many technological components with CHIME. The 256 dish array, represents a total collecting area of $\sim 7200 \mathrm{m^2}$ being built.
It will be located at the South African Radio Astronomy Observatory in the Karoo desert, a few kilometers away from 
the MeerKat site (The South African SKA precursor).   

CHORD ( Canadian Hydrogen Observatory and Radio transient Detector) \cite{2019clrp.2020...28V} can be considered as a successor to CHIME, 
reusing many technologies developed for its predecessor, specially electronics and the FX correlator. But unlike CHIME, it will use 512 dishes,
each $6 \mathrm{m}$ in diameter for the core array, representing a total collecting area of $14400 \mathrm{m^2}$. The bandwidth will be 
slightly increased, covering the 300-1500 MHz  band.
Wide field of view outrigger stations located at large distances from the CHIME/CHORD instruments
will also be built to enhance FRB localisation.  

SKA, a very large general purpose interferometric radiotelescope, is being constructed by an international 
organisation \footnote{SKA observatory: \href{https://www.skaobservatory.org}{https://www.skaobservatory.org} }.
A review of the cosmological surveys planned for SKA Phase-I, 
and their forecasted performance in the redshift range $0<z<3$
can be found in \cite{2020PASA...37....7S}.
In addition to more traditional \HI \, and continuum galaxy surveys at low redshifts $z < 0.35$, Intensity Mapping will enable to extend 
significantly the accessible redshift range, and even cover $3<z<6$ thanks to a deep survey over a $100 \mathrm{deg^2}$ field.  
Finally, one can mention the PUMA white paper \cite{2019arXiv190712559B}, which is a proposal for an ambitious close-packed 
interferometer array with 32000 dishes, covering the frequency range 200-1100 MHz $(0.3<z<6)$, 
as a stage II intensity mapping instrument.   
 
\section{Conclusions, discussion}
\label{sec:conclusion}
21cm Intensity Mapping has emerged in the last decade as a novel, and
possibly powerful method to map large scale structures at redshifts up
to $z \sim 6$. Such surveys would be complementary to optical observations,
and could be used to test and constrain the cosmological model, more
specifically Dark Energy and Inflation. Densely packed interferometers  
with a large bandwidth of a few hundred MHz and large number of receivers 
(several hundreds to several thousands), observing in drift-scan or transit mode, 
without tracking, are the suitable instruments for such surveys. 
These instruments and projects share many technological issues, as well 
data analysis challenges, in particular calibration and foreground subtraction,
with 21cm surveys for EoR, and more broadly, interferometers with large
number of antennae.
Few pathfinder and first stage instruments have been built to explore Intensity Mapping,
and several others are planned or being constructed. 

In the next few years, performance assessment and results from the ongoing 
and future IM experiments will shed light on some of the IM challenges. 
For example, it is important to show that 
it is possible to implement a cost effective design and construction process for a large 
number of antenna and receivers, as well as the associated electronics, while 
maintaining uniformity, construction quality and performance.

Precise array calibration is crucial to achieve large imaging dynamic range and
to ensure foreground subtraction with the required precision.
Among other aspects, it would be necessary to determine beam response for individual feeds,
through a combination of electromagnetic simulations and on site beam measurements.
Arrays with highly redundant baselines present many advantages for the calibration, whereas
such arrays with many identical baselines, perform poorly in terms of mode mixing.

Structures in the instrumental frequency response, due to standing waves for example,
should be minimized and precisely calibrated to be filtered out. Such frequency structures,
if not corrected for, would indeed ruin the foreground subtraction. Another concern
is the contribution of bright sources (sun, radiosources \ldots) through the far side lobes,
which might be highly frequency dependent. It is also important to clarify the level of feed
cross-couplings and correlated noise, which limits instrument sensitivity.

Faraday rotation causes  linearly polarised sources to appear oscillating in frequency.
The two linear polarisation measurements need to be combined during map making,
to remove these oscillations, which in turn implies a good control of
polarisation leakage and calibration. Also, some astrophysical effects
such as self absorbed synchrotron might create higher order mode structures
along the frequency axis, which would impact foreground subtraction. 

Finally, I would like to mention that Line Intensity Mapping (LIM), using other atomic and molecular
lines (CO, Carbon CII \ldots) \footnote{Line Intensity Mapping :
  \href{https://lambda.gsfc.nasa.gov/product/expt/lim\_experiments.html}
  {https://lambda.gsfc.nasa.gov/product/expt/lim\_experiments.html}} 
is also considered for astrophysics and cosmology surveys \cite{2017arXiv170909066K},
although a discussion of LIM is well beyond the scope of this review.

\section*{Acknowledgments}
I would like to thank my Tianlai and PAON4 collaborators, as well as CHIME colleagues,
specially Olivier Perdrereau and Richard Shaw for valuable discussions.

\section*{References}
{\small 
\bibliography{refs} 
}

\end{document}